# Impacts of Stochastic Modeling on GPS-Derived ZTD Estimations

Shuanggen Jin and Jinling Wang
*School of Surveying and Spatial Information Systems*
*The University of New South Wales, Sydney, NSW 2052, Australia*

## BIOGRAPHY

**Shuanggen Jin** is an assistant professor of the Shanghai Astronomical Observatory, Chinese Academy of Sciences. In February 2004 he commenced as a Visiting Fellow in the School of Surveying & Spatial Information System at the University of New South Wales. He received a PhD in Astro-geodynamics and geodesy from the Chinese Academy of Sciences in 2003. His current research is concentrated on GPS data processing and GPS applications in geodynamics and atmospheric sciences. E-mail: sg.jin@unsw.edu.au

**Jinling Wang** is a senior lecturer in the School of Surveying & Spatial Information System at the University of New South Wales. He is Editor-in-Chief of the Journal of Global Positioning Systems, and Chairman of the international study group on pseudolite applications in positioning and navigation within the International Association of Geodesy's Commission 4.

## ABSTRACT

GPS-derived ZTD (Zenith Tropospheric Delay) plays a key role in near real-time weather forecasting, especially in improving the precision of Numerical Weather Prediction (NWP) models. The ZTD is usually estimated using the first-order Gauss-Markov process with a fairly large correlation, and under the assumption that all the GPS measurements, carrier phases or pseudo-ranges, have the same accuracy. However, these assumptions are unrealistic.

This paper aims to investigate the impact of several stochastic modeling methods on GPS-derived ZTD estimations using Australian IGS data. The results show that the accuracy of GPS-derived ZTD can be improved using a suitable stochastic model for the GPS measurements. The stochastic model using satellite elevation angle-based cosine function is better than other investigated stochastic models.

It is noted that, when different stochastic modeling strategies are used, the variations in estimated ZTD can reach as much as 1cm. This improvement of ZTD estimation is certainly critical for reliable NWP and other tropospheric delay corrections.

## INTRODUCTION

The GPS signal propagating through the neutral atmosphere is delayed by variation of the refraction index due to temperature, pressure and water content, which results in lengthening of the raypath, usually referred to as the “tropospheric delay”. This delay is an important error source for GPS positioning. This error source contributes a bias in height of several centimeters even when simultaneously recorded meteorological data are used in tropospheric models (Fang et al., 1998; Tregoning et al. 1998).

Today, GPS has widely been used to determine the ray-path tropospheric delay (Van et al., 1993; Fang et al., 1998; Emardson et al., 1998). The corresponding zenith tropospheric delay (ZTD) can be obtained from the ray-path delay through mapping functions (e.g. Niell, 1996), which can be transformed into the precipitable water vapor (PWV) (Bevis et al, 1994; Duan et al, 1996; Tregoning et al. 1998; Manuel et al., 2001). In comparison with traditional techniques, such as radiosondes and WVR (water vapor radiometer), the GPS technique has more advantages, such as its low cost, all-weather operability and high accuracy (Vedel et al., 2001; Manuel et al., 2001). Therefore GPS-derived ZTD play a key role in near real-time weather forecasting, especially improving the precision of Numerical Weather Prediction (NWP) models. In addition, the GPS-derived ZTD provides a new high-resolution tool for use in atmospheric sciences. Therefore precise GPS-derived ZTD is valuable and beneficial.

ZTD estimation is traditionally obtained using the least squares (LS) principle. In order to employ the LS method for ZTD estimation, both the functional and stochastic models of GPS measurements need to be properly defined. The functional model, also called the mathematical model, describes the mathematical relationships between the GPS measurements and unknown parameters, e.g. baseline components and ZTD. The stochastic model describes the statistical properties of the measurements, which are mainly defined by an appropriate covariance matrix indicating the uncertainty of, and the correlations between, the measurements (Rizos, 1997; Brunner, 1999). Over the past two decades, the functional models for GPS measurements have been investigated in considerable detail. However, accurate stochastic modeling for the GPS measurements is still both a controversial topic and a difficult task to implement in practice (Wang, 1998; Wang et al., 1998; 2002).

In the current stochastic models for use in estimating the ZTD with GPS, it is usually assumed that all the GPS measurements have the same variance. The time-invariant covariance matrix of the double-differenced (DD) measurements is then constructed using the error propagation law. Such assumptions are unrealistic. As GPS measurement errors are dominated by the systematic errors caused by the multipath, ionosphere, and orbit effects, which are quite different for each satellite. Therefore the measurements obtained from different satellites cannot have the same accuracy due to varying noise levels (e.g. Wang, 1998; Wang et al., 1998; Bona, 2000). Previous studies in precise GPS positioning have shown that unrealistic stochastic models may lead to errors of up to 10-15 millimeters in the height components (Satirapod et al., 2002). Therefore, it is expected that the stochastic model will also have a big impact on the precise ZTD estimates. The ZTD is usually estimated by employing the first-order Gauss-Markov (GM) process for calculating variance-covariance (VCV) with a fairly large correlation, namely the power density of 2 cm/sqrt (hour) (http://kreiz.unice.fr/magic/). Whether the first-order GM process and a fairly large correlation are optimal to calculate the ZTD should be further evaluated.

This paper aims to investigate the impact of stochastic models on GPS-derived ZTD using Australian IGS (International GPS service) data. In the following section the data processing method and strategy are described. Stochastic modeling methods and results are then presented.

## DATA PROCESSING METHOD AND STRATEGY

The ionosphere-free linear combination (LC) equation of double-differenced phase can be expressed as:

$$
\begin{aligned}
LC_{kl}^{ij} &= \frac{1}{f_1^2 - f_2^2}(f_1^2 L1_{kl}^{ij} - f_2^2 L2_{kl}^{ij}) \\
&= \rho_{kl}^{ij} + T_{kl}^{ij} + \frac{\lambda_1}{f_1^2 - f_2^2}(f_2^2 N1_{kl}^{ij} - f_1 f_2 N2_{kl}^{ij}) + \varepsilon_{kl}^{ij} \\
&= \rho_{kl}^{ij} + ZTD_k(t)[(m(z_k^i) - m(z_k^j)) - \\
&\quad ZTD_l(t)[(m(z_l^i) - m(z_l^j)) + N_{kl}^{ij} + \varepsilon_{kl}^{ij}
\end{aligned} \tag{1}
$$

where ZTD is the zenith tropospheric delay, $m$ is the mapping function, such as Niell's mapping function (Niell, 1996), $f$ is the frequency, $\overline{N}$ is the (non-integer) ambiguity parameter, and $\varepsilon$ is noise. And the linear observation equations are expressed as:

$$L = Ax + v \tag{2}$$

In Eq. (2), A is the design matrix, L is a vector of the observed-minus-computed DD carrier phase values (O-C), and x is the unknown parameters including ZTD, baseline and ambiguities.

The stochastic models used in ZTD estimation are: a) the covariance matrix for the double-differenced GPS measurements, and b) the stochastic properties of the ZTD parameters which are usually described by the Gauss-Markov process. In some cases the baseline components may also be stochastically constrained. The stochastic models for GPS measurements can be estimated by the MINQUE method (Wang, 1998; Wang et al. 2002) or may be defined by a variety of empirical formulae, whilst the Gauss-Markov process could be set up as the first or second order. In our initial studies, the models to be considered here are the ones that can be easily implemented within scientific GPS data processing software packages such as GAMIT and BERNESE. The details of the variations in the stochastic models are discussed below.

### *Standard stochastic model*

In a commonly-used stochastic model, it's usually assumed that all the carrier phases or pseudo-ranges have the same variance ($\sigma^2$) and are statistically independent. Therefore, the observations $\Phi$ are treated as independent and uncorrelated, and the covariance matrix of the observations $\Phi$ can be formulated as:

$$\mathrm{Cov}(\Phi) = \sigma^2 I \tag{3}$$

where $I$ is the unit matrix. Through the error propagation law the time-invariant covariance matrix (called the stochastic model) of the DD measurements can be obtained:

$$C_x = \sigma^2 \begin{bmatrix} 4 & 2 & \dots & 2 \\ 2 & 4 & \dots & 2 \\ \vdots & \vdots & \ddots & 2 \\ 2 & 2 & \dots & 4 \end{bmatrix} \quad (4)$$

This is a standard stochastic model for DD measurements, which is easy to implement in practice. However, this simplified stochastic model may contain some misspecifications, and thus could result in unreliable ZTD estimates.

***Baseline length dependent weighting***

As the distances between GPS stations in a network are different, the baseline length dependent variances for GPS measurements are defined with the following function (King and Bock, 1999):

$$\sigma^2 = \alpha^2 + \beta^2 * Distance^2 \quad (5)$$

where $\alpha$ = 9mm and $\beta$ = 0.1mm/km. This formula describes the relative qualities of GPS measurements from different GPS baselines in a network.

***Satellite elevation angle dependent weighting***

GPS measurement errors are dominated by systematic errors, such as signal-to-noise ratio, atmospheric delay and multipath errors, which may be closely connected with the satellite elevation angles. The effects of these error sources are different for each satellite. Therefore GPS measurements from different satellites may not have the same accuracy. In order to model the variances of GPS measurements from different satellites, a function of satellite elevation angle is used to describe the variances of raw GPS measurements in practice, namely:

$$\sigma^2{}_{\phi_r^s(i)} = a^2 + b^2 / f^2(elev_r^j(i)) \quad (6)$$

where $a$ and $b$ are constant values, and $f(elev_r^j(i))$ is the function of satellite elevation angle at epoch *i*. Given the variances of one-way GPS measurements, the covariance matrix for the DD measurements is derived using the error propagation law:

$$C_{L(i)} = a^2 \cdot T_{ai} + b^2 \cdot T_{bi} \quad (7)$$

where $T_{ai}$ is the same as Eq. (4), and

$$T_{bi} = \begin{bmatrix} f_{1i} + f_{2i} & f_{1i} & \cdots & f_{1i} \\ f_{1i} & f_{1i} + f_{3i} & \cdots & f_{1i} \\ \vdots & \vdots & \ddots & \vdots \\ f_{1i} & f_{1i} & \cdots & f_{1i} + f_{ni} \end{bmatrix} \quad (8)$$

and $f_{ji} = f(elev_1^j(i)) + f(elev_2^j(i)), j = 1,2,...,n$.

Because of the complexity of unknown factors in the stochastic modelling, the functional relationship between the accuracy of GPS measurements and satellite elevation angles can only be approximately expressed. The sine and cosine functions of satellite elevation angles are often used for this purpose. In the GAMIT software package the sine of the satellite elevation angle is currently used to calculate the accuracy of the one-way GPS measurements (King and Bock, 1999):

$$\sigma^2{}_{\phi_r^s(i)} = a^2 + b^2 / \sin^2(elev_r^j(i)) \quad (9)$$

where $a$ =4.3mm and $b$ =3mm. The function is a good approximation to the tropospheric sensitivity. In addition, the BERNESE software takes another relationship based on satellite elevation angle with the accuracy of the one-way GPS measurements, the cosine of the satellite elevation angle, namely (Hugentobler et al., 2001):

$$\sigma^2{}_{\phi_r^s(i)} = a^2 + b^2 \cdot \cos^2(elev_r^j(i)) \quad (10)$$

For this study, the GAMIT software package has been modified to include this function as one of the options for stochastic modeling.

***Gauss-Markov process***

The GAMIT software parameterizes ZTD as a stochastic variation from the Saastamoinen model with piecewise linear interpolation (King and Bock, 1999). The variation is currently constrained to be a first-order Gauss-Markov process with a special power density (known as the "zenith parameter constraint") of 2 cm/sqrt (hour).

Gauss-Markov (GM) random processes are stationary processes that have exponential autocorrelation functions (Brown and Hwang, 1992). The current GAMIT software package employs a first-order GM process with a fairly large correlation to calculate the variance-covariance (VCV) for ZTD parameter. The autocorrelation function for a second-order GM process X (t) is defined (e.g., Brown and Hwang, 1992) as:

$$R_x(t) = \sigma^2 e^{-\beta|t|}(1 + \beta|t|) \quad (11)$$

The impact of the second GM process on ZTD estimation will not discussed here (but will be evaluated in further studies).

In addition, the baseline components may be treated as observations with defined uncertainty to improve the geometry of the network solutions for the ZTD estimation.

## TEST RESULTS

The test data sets used were from the IGS stations in Australia, as shown in Fig. 1. The GAMIT software was used for the data processing with tightly constrained positions of the IGS stations (each coordinate was assigned a standard derivation of 0.002m). The IGS precise orbits were used. Some parameters used in the data processing were: 1) cut-off elevation: 15 degrees; and 2) GPS data sampling interval: 30 seconds.

The ZTD estimates for these stations on Day 151 2004 were obtained at an interval of two hours using the first-order GM model for the ZTD parameters with "zenith parameter constraint" of 2 cm/sqrt(hour). The performance of the following four stochastic modeling methods for GPS measurements was evaluated:

A: Standard GPS processing method with a simplified stochastic model;
B: Baseline length dependent weighting;
C: Satellite elevation angle-based Sine function; and
D: Satellite elevation angle-based Cosine function.

Fig. 2 shows the standard deviations of the ZTD estimates at the tid1 station using different stochastic models. The differences between the standard deviations for Methods A and D are the biggest ones, reaching 2mm, while the corresponding differences between the ZTD estimates is up to 1cm (Fig. 3).

In addition, the standard deviations of the ZTD estimates based on the baseline length weighting (Method B) is higher than those from the standard processing method (A). The satellite elevation angle-based models (Method C and D) are better than other tested stochastic models. Among all the tested stochastic modeling methods, Method D has the best performance.

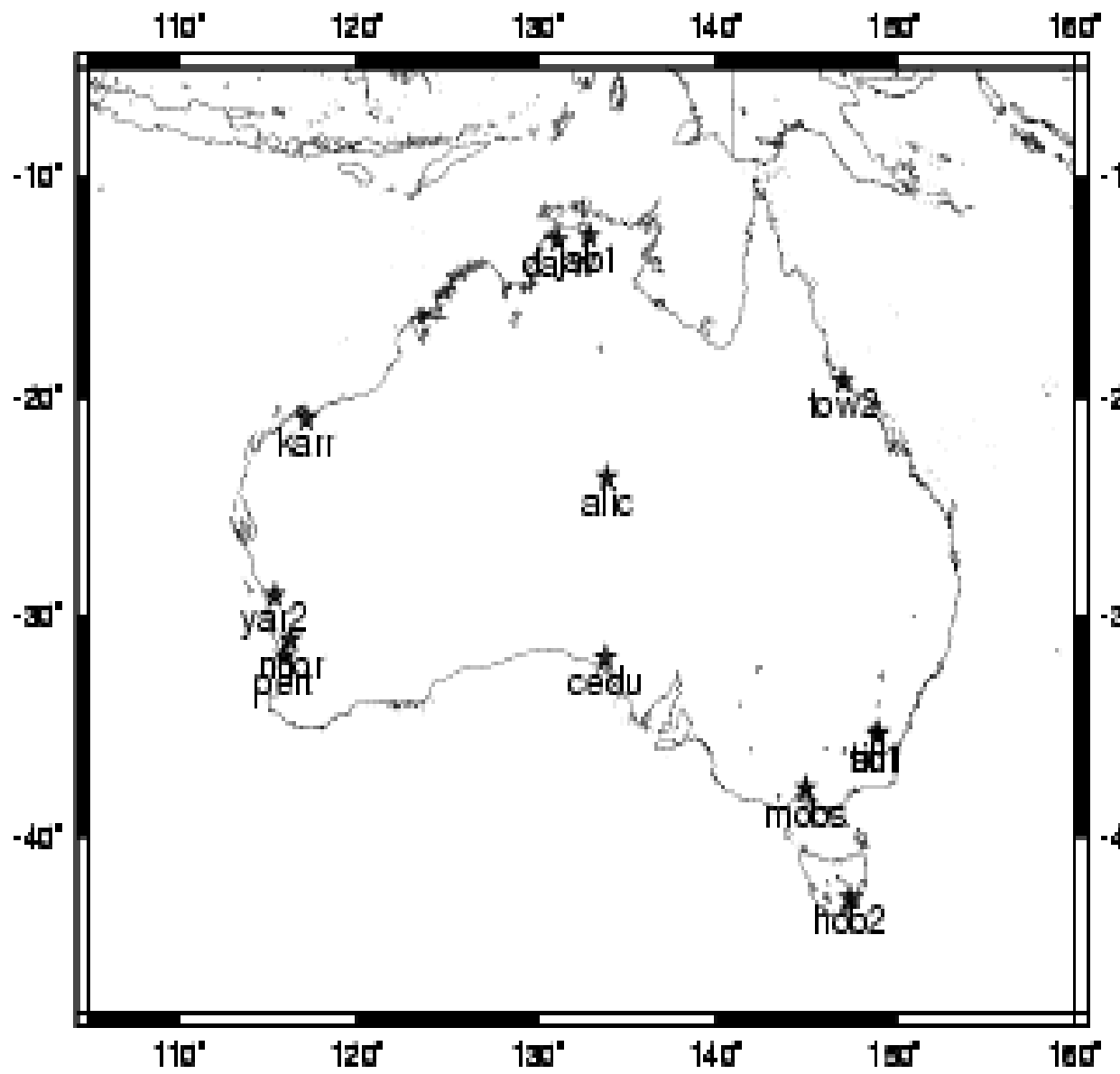

Figure 1 The distribution of GPS stations in Australia

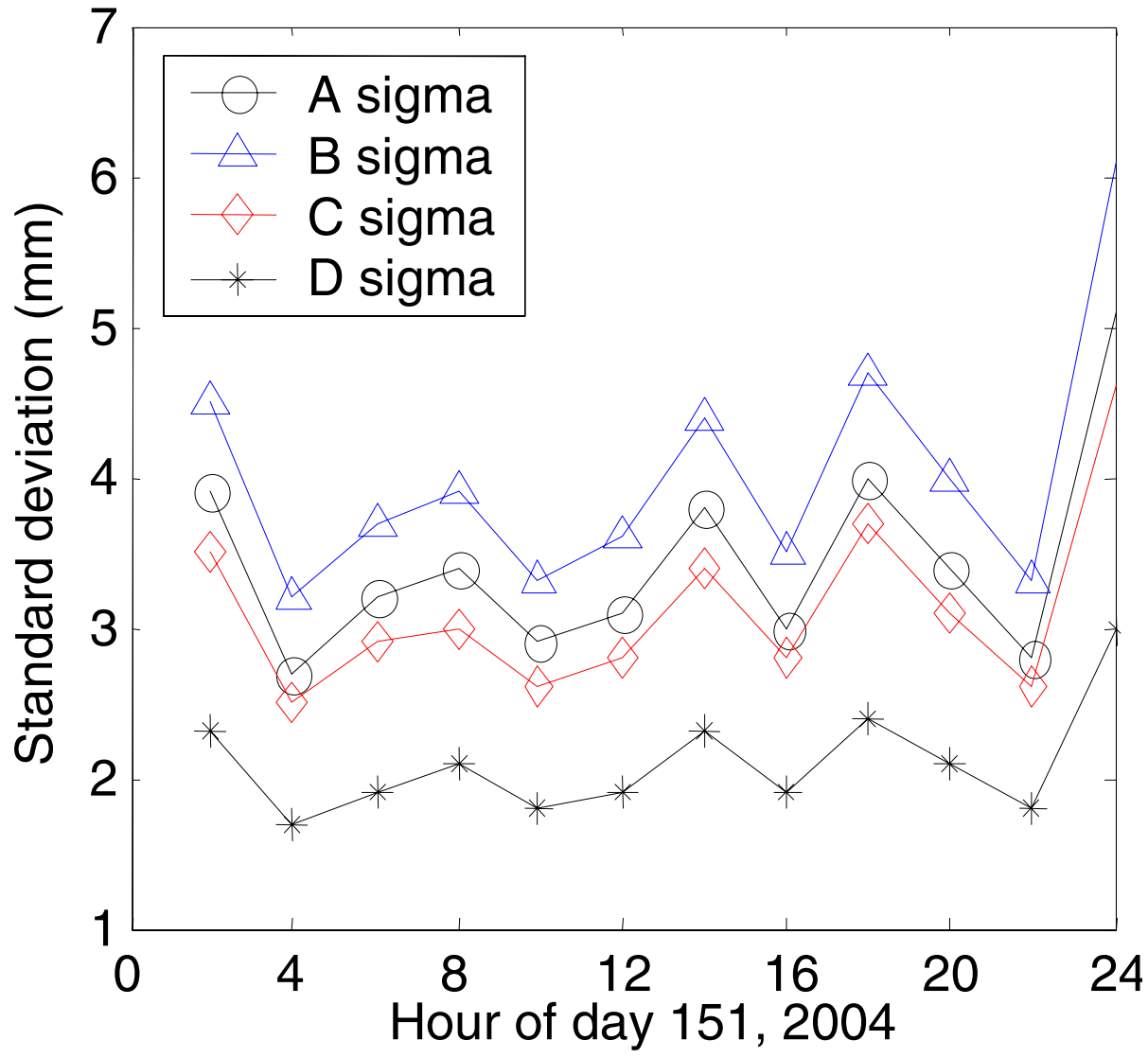

Figure 2 Standard deviations of ZTD estimates at tid1 station using different stochastic models

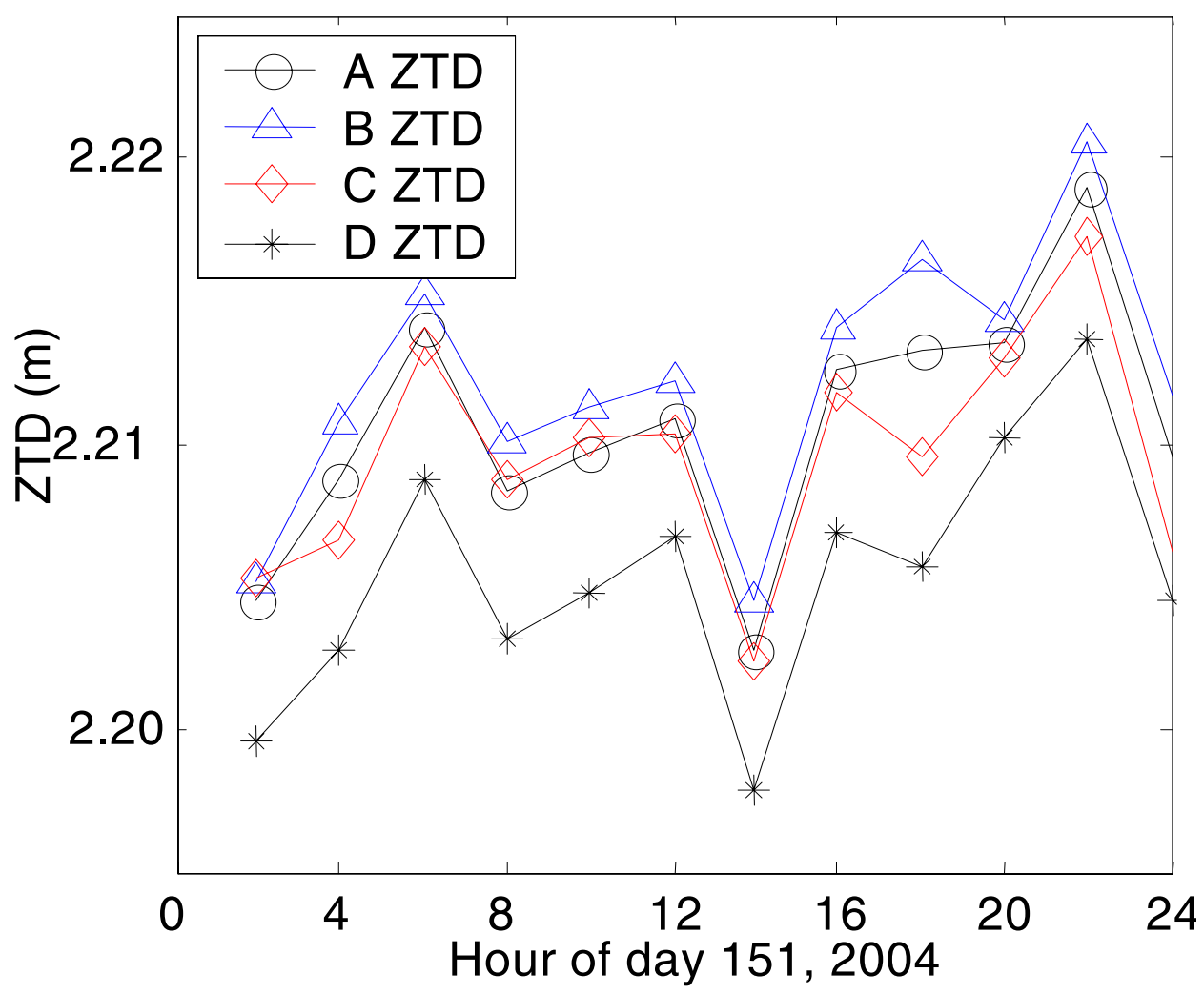

Figure.3 ZTD estimates at the tid1 station using different stochastic models.

The standard deviations and ZTD estimates from the hob2 station are listed in Tables 1 and 2 respectively. The averaged standard deviation for Method D is 2mm smaller than Method B. The results show that Method D yields more reliable ZTD estimates than the other tested methods. In the interval 17-19, the difference between the ZTD estimates from Methods B and D is up to 1.2cm.

Table.1 The standard deviations of the ZTD estimates at the hob2 station

| Intervals (hour) | Standard deviations (mm) | | | |
|---|---|---|---|---|
| | A | B | C | D |
| 1-3 | 4.1 | 4.9 | 3.6 | 2.5 |
| 3-5 | 2.9 | 3.4 | 2.7 | 2.0 |
| 5-7 | 3.5 | 4.2 | 3.2 | 2.2 |
| 7-9 | 3.8 | 4.6 | 3.5 | 2.4 |
| 9-11 | 3.3 | 3.9 | 3.0 | 2.1 |
| 11-13 | 3.4 | 4.0 | 3.1 | 2.2 |
| 13-15 | 4.0 | 4.8 | 3.5 | 2.5 |
| 15-17 | 3.2 | 3.8 | 2.9 | 2.1 |
| 17-19 | 4.5 | 5.4 | 4.3 | 2.8 |
| 19-21 | 4.0 | 4.8 | 3.7 | 2.5 |
| 21-23 | 3.1 | 3.7 | 2.9 | 2.0 |

The above test results show that misspecifications in the stochastic models will result in unreliable ZTD estimation. Using Method D the precisions of GPS-derived ZTD can be improved.

Table.2 The estimated ZTD values at the hob2 station

| Intervals (hour) | ZTD estimations (m) | | | |
|---|---|---|---|---|
| | A | B | C | D |
| 1-3 | 2.397 | 2.398 | 2.396 | 2.391 |
| 3-5 | 2.391 | 2.393 | 2.392 | 2.385 |
| 5-7 | 2.399 | 2.400 | 2.398 | 2.392 |
| 7-9 | 2.427 | 2.428 | 2.429 | 2.422 |
| 9-11 | 2.426 | 2.427 | 2.423 | 2.420 |
| 11-13 | 2.416 | 2.417 | 2.419 | 2.412 |
| 13-15 | 2.399 | 2.401 | 2.398 | 2.393 |
| 15-17 | 2.402 | 2.403 | 2.401 | 2.396 |
| 17-19 | 2.382 | 2.386 | 2.380 | 2.374 |
| 19-21 | 2.389 | 2.389 | 2.387 | 2.384 |
| 21-23 | 2.408 | 2.409 | 2.407 | 2.402 |

## CONCLUDING REMARTKS

The ZTD was usually estimated using the first-order Gauss-Markov process and under the assumption that all the GPS measurements have the same variance.

It has been noted that, with different stochastic modeling methods tested in the paper, the changes in the estimated ZTD values can reach 1.2cm, which is significant for some ZTD applications.

The stochastic modeling testing results here have shown that misspecification in the stochastic models will result in unreliable ZTD estimations. Using the satellite elevation angle-based cosine function the precision of GPS-derived ZTD estimations can be improved.

This improvement of GPS-derived ZTD is certainly critical for reliable numerical weather prediction applications and other tropospheric research. This initial study has demonstrated that the stochastic model methods play an important role in the ZTD estimation process. Suitable stochastic modeling strategies for GPS measurements, baseline components (or the coordinates of GPS tracking stations) and the ZTD parameters should be further investigated.

## ACKNOWLEDGMENTS

The authors would like to thank Dr Bob King and Dr Peng Fang for their valuable discussions and help in this study.